\documentclass[number,sort&compress,preview]{elsarticle}
\usepackage{amsmath}
\usepackage{amssymb}
\usepackage[T1]{fontenc}
\usepackage{mathptmx} 
\usepackage[scaled=.90]{helvet}
\usepackage{courier}
\usepackage{miller}

\usepackage{natbib}
\usepackage{xspace}

\usepackage{mathtools}
\usepackage{nicefrac}
\usepackage{multirow}
\usepackage{booktabs}
\usepackage[hypertexnames=true,colorlinks]{hyperref}
\usepackage[usenames,dvipsnames]{xcolor}
\hypersetup{pdffitwindow=true,linkcolor=blue,citecolor=blue,urlcolor=blue}
\usepackage{microtype}
\usepackage{placeins}
\usepackage{textcomp}
\usepackage{multicol}
\usepackage{fancyhdr}
\fancypagestyle{pprintTitle}{\pagestyle{fancy}}
\newcommand{\equ}[1]{Eq.~\ref{#1}}
\newcommand{\fig}[1]{Fig.~\ref{#1}}

\newcommand{\ie}{i.e.\ }

\newcommand*\mean[1]{\left\langle#1\right\rangle} 

\newcommand{\natoms}{\ensuremath{N}\xspace}
\newcommand{\ntypes}{\ensuremath{N_T}\xspace}

\newcommand{\betavark}%
{\ensuremath{\beta(\left\lbrace{w_k}\right\rbrace)}\xspace}
\newcommand{\emm}{\left(m\right)}
\newcommand{\enn}{\left(n\right)}
\usepackage{caption}
\journal{Scripta Materialia}
\begin{document}
\begin{frontmatter}
\title{Correlation of microdistortions with misfit volumes in High Entropy Alloys}
\author[epfl,imtek]{Wolfram Georg N\"ohring \corref{cor1}}
\ead{wolfram.noehring@imtek.uni-freiburg.de}
\cortext[cor1]{Corresponding author}
\author[epfl]{W.\ A.\ Curtin}
\address[epfl]{Institute of Mechanical Engineering, \'Ecole Polytechnique F\'ed\'erale de Lausanne, EPFL STI IGM Station 9, CH-1015 Lausanne}
\address[imtek]{Department of Microsystems Engineering, University of Freiburg, 79110 Freiburg, Germany}
\date{January 14, 2019}
%
\begin{abstract}
The yield strengths of High Entropy Alloys have recently been correlated with measured picometer-scale atomic distortions. Here, the root mean square microdistortion in a multicomponent alloy is shown to be nearly proportional to the misfit-volume parameter that enters into a predictive model of solute strengthening. Analysis of two model ternary alloy families, face-centered cubic Cr-Fe-Ni and body-centered cubic Nb-Mo-V, demonstrates the correlation over a wide composition space. The reported correlation of yield strength with  microdistortion is thus a consequence of the correlation between microdistortion and misfit parameter and the derived dependence of yield strength on the misfit parameter.
\end{abstract}
\begin{keyword}
    random alloys, solute-solute interactions, short-range order, modeling
\end{keyword}
\end{frontmatter}
%
%
High Entropy Alloys (HEA) are a new class of crystalline alloys, consisting of many elemental components at non-dilute concentrations randomly distributed among the crystalline lattice sites.  Some of these HEAs show impressive mechanical properties --- strength  ~\cite{senkov_refractory_2010,senkov_mechanical_2011}, ductility ~\cite{wu_temperature_2014,wu_recovery_2014,li_metastable_2016,yao_novel_2014}, fracture toughness \cite{gludovatz_fracture-resistant_2014}, and enhanced resistance to hydrogen embrittlement ~\cite{luo_hydrogen_2017,nygren_hydrogen_2018,luo_corrosion_2018}.  These findings are driving experimental and theoretical studies into the mechanistic origins of the mechanical properties of HEAs ~\cite{miracle_critical_2017}.  In particular, various works have shown correlations of yield strength (at finite temperature, or extrapolated to $T=0$~K) with atomic misfit volumes ~\cite{yao_nbtav-tiw_2016,toda-caraballo_interatomic_2015,yao_mechanical_2017}, picometer-scale microdistortions \cite{okamoto_atomic_2016,oh_lattice_2016,wang_impacts_2017,chen_contribution_2018}, and electronic structure.  The physical origins of these correlations remain unclear, but can be used for ad-hoc selection of new possible alloys.   

Of particular interest here is the correlation of strength with microdistortions.  Experimental measurements of the deviations of atomic positions from the exact lattice positions, after subtracting the thermal contributions, have been characterized by the root mean square distortion 
\begin{align}
    u_\mathrm{rms} = \sqrt{\left\langle\sum_{i \in \text{sites}} \left\vert\mathbf{r}_{i}-\mathbf{r}_{i0}\right\vert^2\right\rangle}, \label{equ:msd}
\end{align}
where $\mathbf{r}_i$ and $\mathbf{r}_{i0}$ are the position of site $i$ in the distorted and 
perfect lattice, respectively, and $\left\langle\dots\right\rangle$ indicates the average with
respect to site occupations. 
\newcommand{\mualloy}{\ensuremath{\mu_\mathrm{alloy}}}
Measured yield strengths in the Co-Cr-Fe-Mn-Ni family of alloys have been extrapolated to $T=0$~K \cite{wu_temperature_2014}, normalized by the measured alloy shear modulus $\mu_\mathrm{alloy}$, and then correlated with the MSD as 
\begin{align}
\sigma_{y0}/\mualloy \propto u_{rms}.  \label{equ:correlation}
\end{align}
Other recent work ~\cite{wang_impacts_2017,chen_contribution_2018} is using microdistortions computed using first-principles density functional theory to rationalize relative strengths in various BCC HEAs.  There is a general physical concept that the dislocation interacts with those microdistortions in some manner, but the connection is not precise.  The strengths in HEAs also have a contribution due to grain size (\ie the Hall-Petch effect) that is not directly connected to the atomic scale motion of dislocations through the random alloy.  This effect should be subtracted out from raw experimental data to reveal the ``chemical'' effect of the random alloy.  Some of the above correlations have been made using the total strength, and this leads to some uncertainty in the results.

Here, we show that there is a good correlation between the MSD and the misfit parameter that enters a predictive solute strengthening model for the chemical contribution to strengthening (see below).  This fully rationalizes the correlation of \equ{equ:correlation}.  There is thus no hidden physics or mechanics in  \equ{equ:correlation} --- the correlation rather serves to demonstrate that the ``chemical'' strengthening is ultimately connected to the interaction energy between the dislocation and the solute misfit volumes in the random alloy, as captured by the predictive theory.

The full predictive theory for the chemical component of the strengthening in fcc alloys connects the temperature and strain-rate dependent yield stress to underlying solute/dislocation interaction energies and other fundamental properties of the alloy ~\cite{varvenne_theory_2016}.  The theory envisions an alloy with $n=1,2\dots\ntypes $ components at concentrations ${c_n}$ distributed randomly on the crystalline lattice.  When simplified to elasticity theory ~\cite{varvenne_theory_2016}, the theory predicts that the zero-temperature yield strength and energy barrier for thermal activation scale as 
\begin{align}
    \begin{aligned}
        \tau_{y0}           & \propto \mualloy \left(\frac{1+\nu}{1-\nu}\right)^{\frac{4}{3}} \left[\frac{1}{b^6}\sum_{n} c_n 
        \Delta V_n^2 \right]^\frac{2}{3},\\
        \quad  \Delta E_{b} & \propto \mualloy \left(\frac{1+\nu}{1-\nu}\right)^{\frac{2}{3}} \left[\frac{1}{b^6}\sum_{n} c_n 
        \Delta V_n^2 \right]^\frac{1}{3},
    \end{aligned} \label{equ:volumescaling}
\end{align}
leading to the yield strength
\begin{align}
    \tau_y \left(T, \dot{\varepsilon}\right) = \tau_{y0}   \left[1-\left(\frac{kT}{\Delta E_b}\ln{\frac{\dot{\varepsilon}_0}{\dot{\varepsilon}}}\right)^{\frac{2}{3}}\right]
\end{align}
as a function of temperature T and strain rate $\dot{\varepsilon}$.  In  \equ{equ:volumescaling}, $\Delta V_n$ is the average misfit 
 volume of the $n$th component in the alloy and $b$ is the magnitude of the dislocation Burgers vector.  The strength thus depends on the misfit parameter $\delta '$ given by 
 \begin{align}
 \delta ' = \frac{1}{b^3}\sqrt{\sum_{n} c_n 
        \Delta V_n^2},
\end{align}
which is closely related to the traditional $\delta$ parameter defined as
\begin{align}
    \delta = \frac{1}{3V_{alloy}} \sqrt{\sum_{n} c_n \Delta V_n^2}, 
\end{align}
where $V_{alloy}$ is the average atomic volume of the alloy ($V_{alloy} = b^3/\sqrt{2}$ for fcc and $V_{alloy} = 3 \sqrt{3} b^3 /8$ for bcc).  The measured tensile yield strength $\sigma_{y}$ is obtained from $\tau_y$ as $\sigma_y = M \tau_y$ where, for instance, $M=3.06$ is the Taylor factor for an untextured fcc polycrystal.  Neglecting the term involving the Poisson ratio, the zero-temperature yield strength is thus predicted to scale as
\begin{align}
    \sigma_{y0} / \mualloy \propto \delta'^{4/3}. 
\end{align}
Therefore, if the RMSD $u_\mathrm{rms}$ is correlated to the misfit parameter $\delta'^{4/3}$, then the observed correlation between yield strength and RMSD is understood fundamentally as due to the underlying solute misfit volumes in the alloy. We therefore now show that such a correlation exists.

\begin{figure}[htb!]
    \centering
    \includegraphics{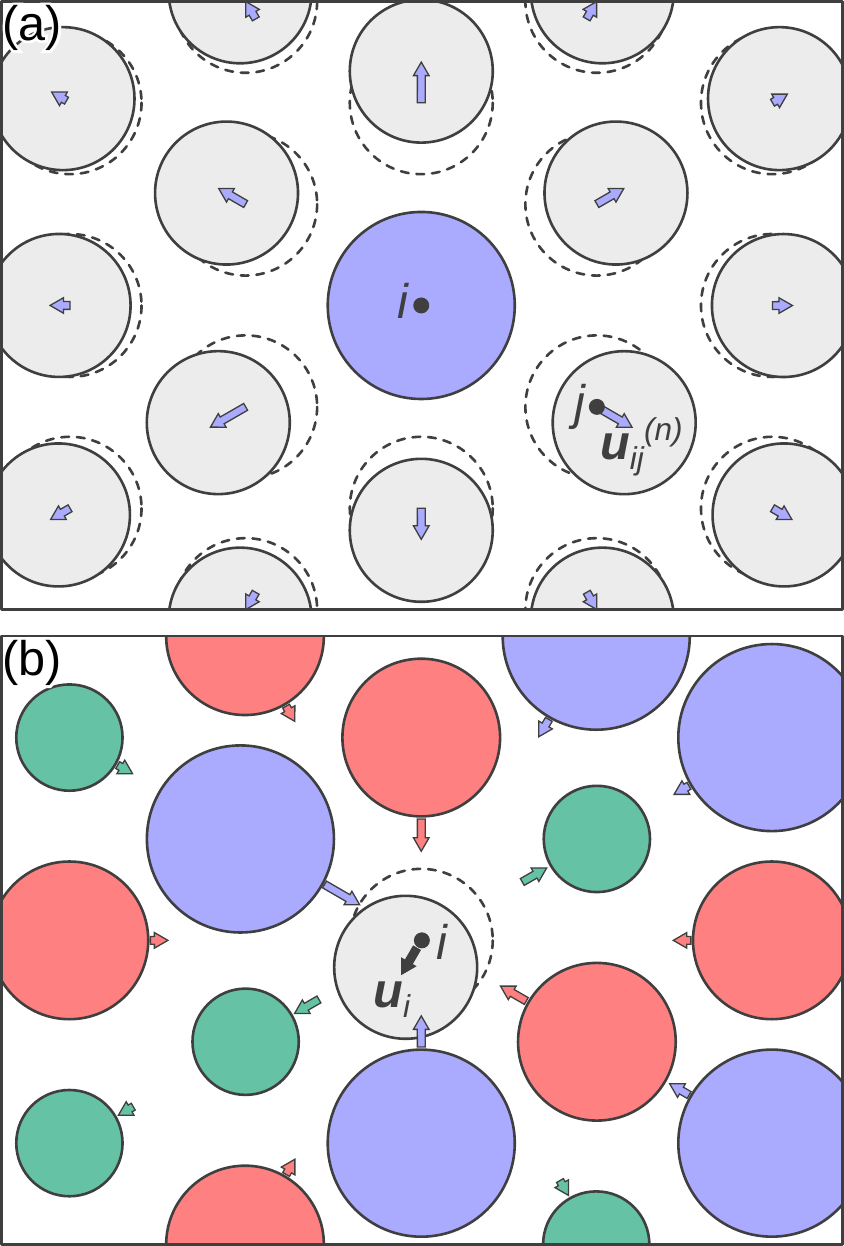}
    \caption{(a) Displacement field created by a single solute in the average matrix; (b) Displacement of an atom at position $i$ created by the sum of the displacements at site $i$ generated by all of the surrounding solutes on sites ${j}$.  For simplicity, the schematic only shows displacements out to second neighbors and is not to scale.}
    \label{fig:displacementvectors}
\end{figure}

We postulate that the microdistortion at a given atomic site $i$ in the lattice is determined by the atomic displacements $\mathbf{u}_{ij}$ due to the misfits of the atoms at all other sites ${j}$ in the vicinity of site $i$.  The sum of the vector displacements contributed by all sites ${j}$ and over all solute types ${n}$ at concentrations ${c_n}$ is exactly zero; \ie the average alloy is a perfect crystalline lattice.  However, the fluctuations due to the random arrangement of specific solutes around any given lattice site leads to a RSMD at that site.  This RMSD is computed as the root-mean-square displacements generated by the misfit distortions of all the surrounding atoms on the central atom.  \fig{fig:displacementvectors} shows a schematic diagram of this concept, with \fig{fig:displacementvectors}a) illustrating the average displacements cause on surrounding atoms by one solute type, and \fig{fig:displacementvectors}b) illustrating the net displacement on the central atom due to the displacements caused by the specific arrangement of surrounding solute atoms.

The RMSD is computed from the atomic 
displacement fields explicitly as follows. Let 
$u_{i,p}^{(m)}$ be the $p$-component of the displacement vector generated at site $i$ by a solute of type $m$ located at the origin.  This displacement is the average displacement when the solute is introduced into the average matrix.  The random environment is described by the occupation variables ${s_{i}^{(m)}}$ where $s_{i}^{(m)}=1$ if there is a type $m$ atom at site $i$ is of type $m$ and zero otherwise. The root mean square of the displacement field on a central atom is computed in terms of all surrounding atoms at positions ${i}$ as
\begin{align}
    u_\mathrm{rms} = \sqrt{\sum_p^3 \mean{\left(\sum_{i}^N \sum_n^{\ntypes} s_i^{(m)} u_{i,p}^{(m)}\right)^2}}. \label{equ:a1}
\end{align}

Evaluation of the RMSD is accomplished first by expanding the square and identifying four partial sums over disjoint index sets as
\begin{align}
\begin{aligned}
\mean{\left(\sum_{i}^N \sum_n^{\ntypes} s_i^{(m)} u_{i,p}^{(m)}\right)^2} &= 
\mean{
	\sum_{m}^{\ntypes}
	\sum_{n}^{\ntypes}
	\sum_{i}^{\natoms}
	\sum_{j}^{\natoms}
	s_{i,p}^{\left(m\right)}
	s_{j,p}^{\left(n\right)}	
	u_{i,p}^{\left(m\right)}
	u_{j,p}^{\left(n\right)}
} 
\\ &= \mean{\Sigma_1} + \mean{\Sigma_2} + \mean{\Sigma_3} + \mean{\Sigma_4},
\end{aligned}
\label{equ:a2}
\end{align}
where
\begin{align}
\begin{aligned}
\Sigma_1&=
\sum_{m}^{\ntypes}
\sum_{i}^{\natoms}
\left(s_{i}^{\left(m\right)}	
u_{i,p}^{\left(m\right)}\right)^2, \\
\Sigma_2 &= 
\sum_{m}^{\ntypes}
\sum_{i}^{\natoms}
\sum_{j \neq i}^{\natoms}
s_{i}^{\left(m\right)}
s_{j}^{\left(m\right)}	
u_{i,p}^{\left(m\right)}
u_{j,p}^{\left(m\right)}, \\
\Sigma_3 &=
\sum_{m}^{\ntypes}
\sum_{n \neq m}^{\ntypes}
\sum_{i}^{\natoms}
s_{i}^{\left(m\right)}
s_{i}^{\left(n\right)}	
u_{i,p}^{\left(m\right)}
u_{i,p}^{\left(n\right)},\\
\Sigma_4 &= 
\sum_{m}^{\ntypes}
\sum_{n \neq m}^{\ntypes}
\sum_{i}^{\natoms}
\sum_{j \neq i}^{\natoms}
s_{i}^{\left(m\right)}
s_{j}^{\left(n\right)}	
u_{i,p}^{\left(m\right)}
u_{j,p}^{\left(n\right)}.
\end{aligned}
\end{align}
$\Sigma_3$ is zero because the 
same site $i$ cannot be occupied by different
types $m$ and $n$.  
The occupation variables are Bernoulli random variables with
$ \langle s_{i}^{\emm} \rangle =\langle (s_{i}^{\emm})^2\rangle=c_m$. 
In the random alloy, the occupation variables at different sites are
uncorrelated so \ie $\langle s_{i}^{\emm}s_{j}^{\enn}\rangle=c_mc_n$.
The averages of the partial sums therefore become 
\begin{align}
\begin{aligned}
\mean{\Sigma_1} &=
\sum_{m}^{\ntypes}
\sum_{i}^{\natoms}
c_m\left(u_{i,p}^{\emm}\right)^2,\\
\mean{\Sigma_2} &= 
\sum_{m}^{\ntypes}
\sum_{i}^{\natoms}
\sum_{j \neq i}^{\natoms} c_m^2 u_{i,p}^{\emm}  u_{j,p}^{\emm}, \\ 
\mean{\Sigma_4} &=
\sum_{m}^{\ntypes}
\sum_{n \neq m}^{\ntypes}
\sum_{i}^{\natoms}
\sum_{j \neq i}^{\natoms}
c_mc_n u_{i,p}^{\emm} u_{j,p}^{\enn}. \\
\end{aligned}
\label{equ:a4}
\end{align}
Since the net displacement is zero due to inversion symmetry
about the origin, \ie 
\begin{align}
\sum_{i}^{\natoms} u_{i,p}^{\emm} = 0,
\end{align}
we have after some rearrangement
\begin{align}
\begin{aligned}
\mean{\Sigma_2} &= 
-\sum_{m}^{\ntypes}c_m^2
\sum_{i}^{\natoms} \left(u_{i,p}^{\emm}\right)^2,				
\end{aligned} 
\label{equ:a6}
\end{align}
and
\begin{align}
\begin{aligned}
\mean{\Sigma_4} &=
-\sum_{m}^{\ntypes}c_m
\sum_{n \neq m}^{\ntypes}c_n 
\sum_{i}^{\natoms}
u_{i,p}^{\emm} u_{i,p}^{\enn}.
\end{aligned}
\label{equ:a7}
\end{align}
Combining \equ{equ:a6}, \equ{equ:a7}, and $\mean{\Sigma_1}$ from \equ{equ:a4} gives the following result for \equ{equ:a1},
\begin{align}
u_\mathrm{rms}= \sqrt{\sum_p^3 \left[\sum_{m}^{\ntypes}c_m
	\sum_{i}^{\natoms}
	\left(u_{i,p}^{\emm}\right)^2 - \sum_{m}^{\ntypes}c_m
	\sum_{n}^{\ntypes}c_n 
	\sum_{i}^{\natoms}
	u_{i,p}^{\emm} u_{i,p}^{\enn}\right]}.
\end{align}
We neglect the second term, which involves products of concentrations and products of displacements that are both positive and negative.  We normalize the displacements by the Burgers vector $b$ as $u_{i,p}^{\emm} \rightarrow b \tilde{u}_{i}^{\emm}$. This lead to our final result for the normalized RMSD
\begin{align}
u_\mathrm{rms}/b \approx \sqrt{\sum_p^3 \sum_{m}^{\ntypes}c_m
	\sum_{i}^{\natoms}
	\left(\tilde{u}_{i,p}^{\emm}\right)^2 }. \label{equ:rmsdapprox}
\end{align}
In \equ{equ:rmsdapprox}, the normalized RMSD is fully determined by the normalized misfit-induced atomic displacements and the alloy composition.  We must now demonstrate that this misfit-induced RMSD is related to the $\delta '$ parameter that controls alloy strengthening.  

The correlation is demonstrated using atomistic simulations for alloys described by a set of EAM-type interatomic potentials.  We first construct the ''average alloy" potential that encodes all the average properties of that alloy into a single average-atom EAM-type potential \cite{varvenne_average-atom_2016}.  We then introduce one individual atom of type $m$ into the average alloy and measure the solute misfit volume $\Delta V_m$ and the induced atomic displacement field ${u}_{i,p}^{\emm}$ at all sites $i$ around the solute.  Performing this procedure for all solute types $m$ provides the entire set of displacements needed to evaluate \equ{equ:rmsdapprox} above.  

\begin{figure}[htb!]
    \centering
    \includegraphics{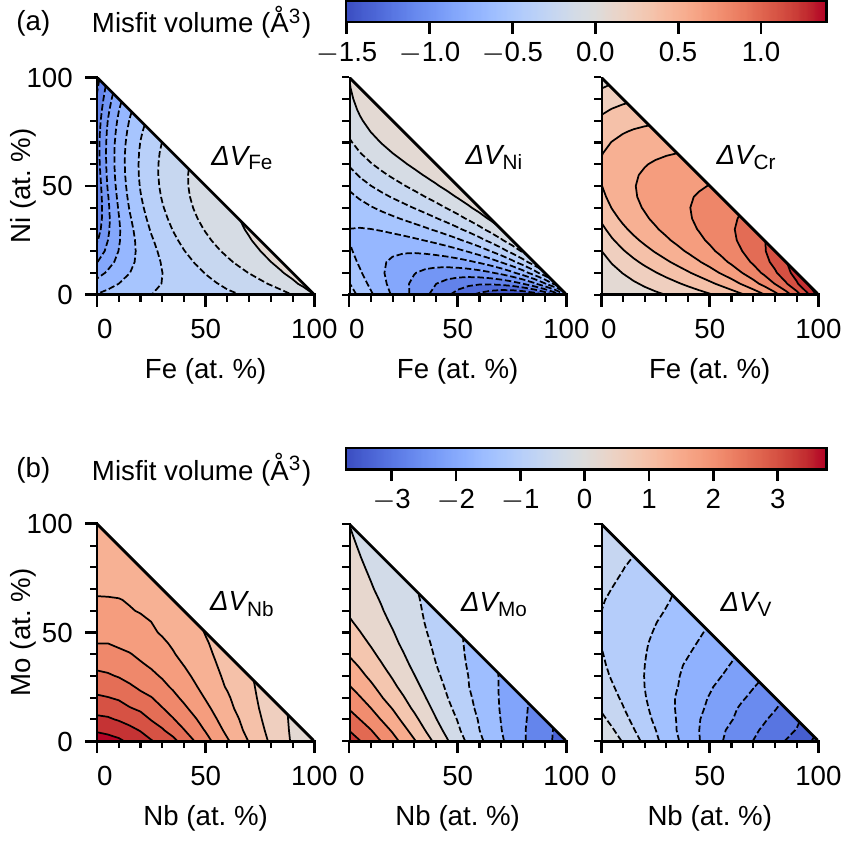}
    \caption{(a) Misfit volumes of Fe, Ni, and Cr solutes in the average Cr-Fe-Ni alloy family; (b)
    (b) Misfit volumes of Nb, Mo, and V solutes in the average Nb-Mo-V alloy family.}
    \label{fig:volumes}
\end{figure}

\begin{figure}[htb!]
    \centering
    \includegraphics{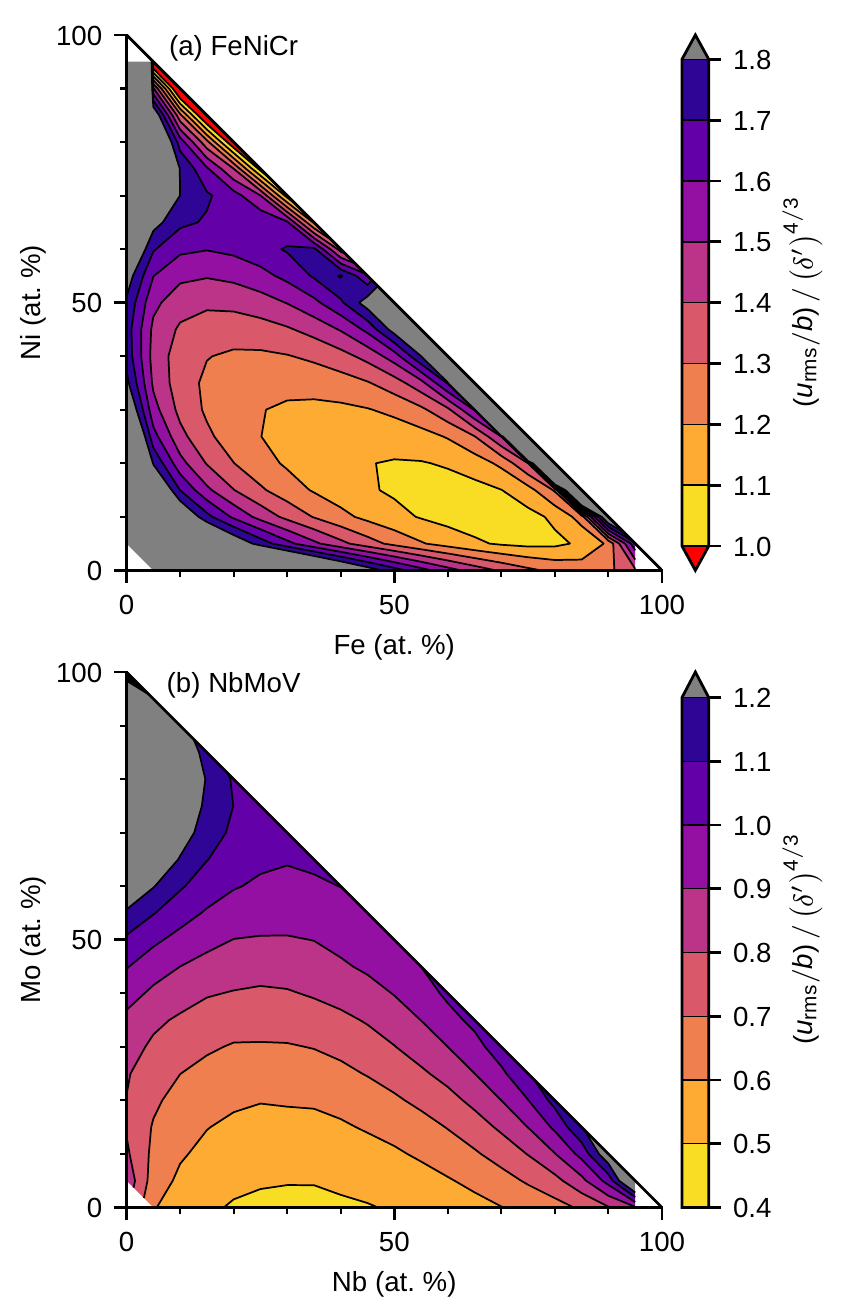}
    \caption{
    Dimensionless ratio $\left(u_\mathrm{rms}/b\right)/\left(\delta'\right)^{4/3}$ for (a) the Cr-Fe-Ni system and 
    (b) the Nb-Mo-V system. 
    \label{fig:ratios}
    }
\end{figure}

We have executed the above strategy for two different families of ternary alloys, the fcc Cr-Fe-Ni system and the bcc Nb-Mo-V system, using available interatomic potentials \cite{bonny_interatomic_2011,zhou_atomic_2001,rao_notitle_nodate}.  Figs.~\ref{fig:volumes}a,b) show the misfit volumes of each solute type as a function of composition for these two alloy families. Figs.~\ref{fig:ratios}a,b)  show the ratio of the normalized RMSD to $\left(\delta'\right)^{4/3}$ across the entire composition range for both alloy families.  Fig.~\ref{fig:volumes}a)  shows that the ratio for fcc Cr-Fe-Ni is $\approx 1.2 \pm 0.2$ over a wide range of compositions in the ``High Entropy'' or non-dilute range for any of the elements.   Fig.~\ref{fig:volumes}b) shows that the ratio for bcc Mo-Nb-V is $0.7 \pm 0.2$ again over a wide range of compositions in the ``High Entropy'' domain.  Results for Nb-Ta-V (not shown), which is like a pseudo-binary (NbTa)-V alloy, are similar but with even less variation.

The good correlation between the RSMD and $\delta '$ in the High Entropy domain demonstrated here shows that the correlation between RSMD and yield strength is primarily an alternative manifestation of the role played by the collective misfit volumes of the alloy in determining the resistance of the material to dislocation motion through the random structure.  Since the experimental measurement of microdistortions is painstaking  \cite{okamoto_atomic_2016}, yet correlated with misfit volumes, it may be more direct to measure misfit volumes through macroscopic experiments or estimate them, e.g. through the use of Vegard's Law.  In a family of alloys such as the Cantor family Co-Cr-Fe-Mn-Ni, the fabrication and measurement of lattice constants for a number of alloys over a range of compositions around a target composition can be used to deduce the individual solute misfit volumes at various compositions by fitting.  The theory also involves the elastic moduli, but these can also be measured on the fabricated materials and trends with composition identified.  With both composition-dependent $\delta '$ and $\mu_{alloy}$  across a range of composition space, the theory can be used to identify composition regimes that should achieve optimal yield strengths.  The composition-dependent $\delta '$ and elastic constants $C_{ij}$ can also be computed using first-principles methods, as recently demonstrated \cite{tehranchi_softening_2017,yin_first-principles-based_nodate,maresca_mechanistic_nodate}.  Vegard's Law and a Rule-of-Mixtures for elastic constants have been shown to be reasonable for estimating the parameters in the theory.  Thus, the need for experimental fabrication and testing can be avoided except for alloys identified by the theory to be most promising.  On the other hand, the RMSD can be easily computed within DFT supercells \cite{wang_impacts_2017,chen_contribution_2018}.  This may be computationally less intensive than computing misfit volumes but (i) the accuracy of the convergence of the microdistortions versus supercell size is uncertain, (ii) the coefficient between RMSD and the misfit quantity may vary from material family to material family, and (iii) the correlation between RMSD and yield strength is only accurate within about $20-25\%$.  Nonetheless, computing microdistortions may provide sufficient guidance for alloy selection. 

Finally, the reduced theory based on misfit volumes and elasticity theory may have quantitative limitations because solute/dislocation interactions in the very core of the dislocation may differ from the elasticity effort, there may be solute-solute interactions, and some HEAs may have short-range order ~\cite{kormann_long-ranged_2017}.  Further theory must thus be developed.  We currently see no path for extension of the microdistortion correlation to address any of these issues.  On the other hand, the theory deviates from experiments at very low temperatures ~\cite{varvenne_theory_2016} and this may be attributable to atomic-scale distortions that are not embedded in the theory.  As explained in Varvennet et al., the theory involves the use of a line tension to capture the dominant solute fluctuations in the random alloy. Atomic distortions, beyond simply the global RMSD, may thus create additional small-scale barriers to dislocation motion that are missing in the theory but that, nonetheless, provide strengthening at low temperatures.  For materials where the low-temperature strength is not of high interest, the current theory and future extensions to include solute-solute interactions and/or short-range order will provide a quantitative framework for prediction of yield strength in HEAs.

%
\section*{Acknowledgment}
Support for  this work  was provided by  the European  Research Council
through the Advanced Grant ``Predictive Computational Metallurgy'', ERC
Grant agreement  No.\ 339081 --  PreCoMet.  Furthermore, support  was
provided by EPFL through the use of the facilities of its Scientific IT
and  Application Support  Center. WGN also acknowledges 
support from the European Research Council through grant ERC-StG-757343.
\appendix
\section*{References}
\bibliographystyle{elsarticle-num}
\bibliography{references.bib}
\end{document}